\documentclass[11pt]{llncs} 
\usepackage{amsmath, amsfonts, amssymb}
\usepackage{hyperref}
\usepackage{url}
\usepackage{amsmath}
\usepackage{graphicx}
\usepackage[margin=1.06in]{geometry}
\usepackage{setspace}
\DeclareGraphicsExtensions{.pdf,.png,.jpg}

\title{...}

\begin{document}
\pagestyle{plain}

\mainmatter
\title{Distributed Gradient Descent in Bacterial Food Search}

\author{Shashank Singh$^{*1}$, Sabrina Rashid$^{*2}$, Zhicheng Long$^{3}$, Saket Navlakha$^4$, Hanna Salman$^5$, Zolt\'{a}n N. Oltvai$^6$, Ziv Bar-Joseph$^{7\dagger}$}
\institute{$^1$ Machine Learning Department and Department of Statistics, Carnegie Mellon University, Pittsburgh, PA 15213.\\
$^2$ Computational Biology Department, Carnegie Mellon University, Pittsburgh, PA 15213.\\
$^3$ Department of Pathology, Department of Physics and Astronomy, University of Pittsburgh, Pittsburgh, PA 15213.\\
$^4$ The Salk Institute for Biological Studies, Center for Integrative Biology, La Jolla, CA 92037.\\
$^5$Department of Physics and Astronomy, Department of Computational and Systems Biology, University of Pittsburgh, Pittsburgh, PA 15260.\\
$^6$Department of Pathology, Department of Computational and Systems Biology, University of Pittsburgh, Pittsburgh, PA 15260.\\
$^7$ Machine Learning Department and Computational Biology
Department, Carnegie Mellon University, Pittsburgh, PA 15213.\\
$^*$ These authors contributed equally.\\
$^\dagger$ Contact author. Email: zivbj@cs.cmu.edu}

\maketitle
\begin{abstract}
Communication and coordination play a major role in the ability of bacterial
cells to adapt to ever changing environments and conditions. Recent work has
shown that such coordination underlies several aspects of bacterial responses
including their ability to develop antibiotic resistance. Here we develop a
new distributed gradient descent method that helps explain how bacterial cells
collectively search for food in harsh environments using extremely limited
communication and computational complexity. This method can also be used for
computational tasks when agents are facing similarly restricted conditions. We
formalize the communication and computation assumptions required for
successful coordination and prove that the method we propose leads to
convergence even when using a dynamically changing interaction network.  The
proposed method improves upon prior models suggested for bacterial foraging
despite making fewer assumptions. Simulation studies and analysis of
experimental data illustrate the ability of the method to explain and further
predict several aspects of bacterial swarm food search.
\end{abstract}
Supporting movies: \url{ https://www.andrew.cmu.edu/user/sabrinar/Bacteria_movies/}
\newpage
\begin{doublespace}
\section*{Introduction}
There are many parallel requirements of computational and biological
systems, suggesting that each can learn from the other. Like
virtually all large-scale computing platforms, biological systems
are mostly distributed consisting of molecules, cells, or organisms
that interact, coordinate, and reach decisions without central
control \cite{Seeley}, \cite{Canright}. However, unlike most
computational methods, biological systems rely on very limited
communication protocols, do not assume that the identity of the
communicating agents is known and only utilize simple computations
\cite{Cheong}. This makes biological systems more resilient to environmental noise and allows them to function efficiently in harsh settings, a property that is desirable in computational systems as
well (for example, sensor networks in remote locations or robot swarms in
mines \cite{Keat}). Recent work demonstrated that for certain network based
distributed algorithms, our improved ability to study biological processes at
the molecular and cellular levels allows us to both, improve our understanding
of these biological processes and suggest novel ways to improve
distributed computational algorithms  \cite{Saket}.

For example, a number of key machine learning optimization algorithms,
including neural networks and non-negative matrix factorization, have been
inspired by information processing in the brain \cite{Hopfield},
\cite{Bishop}. Here we show that a variant of a commonly used
machine learning coordination algorithm, \emph{distributed gradient descent
(DGD)}, can be used to explain how large bacterial swarms efficiently search for food.
Similar to the regular gradient descent setting \cite{Hastie}, \cite{Mzard},
by sensing the food gradient, each cell has its own belief about the location
of the food source (Figure ~\ref{model}a). However, given potential obstacles in the
environment, as well as limits on the ability of each
cell to accurately detect and move toward the food source in this
environment, the individual trajectories may not produce the optimal path to
the food source. Thus, in addition to using their own belief each cell also
sends and receives messages from other cells (either by secreting specific signaling molecules or by physical interaction; Figure ~\ref{model}(b)), which it then integrates to
update its own belief and determine subsequent movement direction and velocity.
The process continues until the swarm converges to the food source.

While it is possible to observe that the group-based approach shortens the time it takes
a single cell to reach the food source (both experiments and simulations
as we show below), and several aspects of the
molecular pathways involved in the communication between bacterial
cells have been studied \cite{Waters}, the {\em computations} that
the cells perform have not been been well characterized. Current
models \cite{torney},\cite{shklarsh} are largely based on
differential equation methods. While these can indeed lead to a fast,
coordinated search, they do not fully take into account the dynamically
changing topology of cells' interaction network over time. Furthermore, they
make unrealistically strong assumptions about the abilities of cells to
identify the source(s) of the messages and to utilize a large (effectively
continuous valued) set of messages, which is unrealistic given the limited
computational powers bacteria cells possess.

Here, we introduce a distributed gradient descent (DGD) model that makes biologically
realistic assumptions regarding the dynamics of the agents, the size of the set
of the messages, and the agents' ability to identify senders. We show that our DGD model solves the bacterial food search problem more efficiently (in terms of the overall complexity of
messages sent) and more quickly (in terms of the time it takes the swarm to
reach the food source) when compared to current differential equation models and also leads to better agreement with experimental results.
We argue that our model is in fact a distributed pseudogradient descent method,
\footnote{\emph{pseudogradient} here refers to the fact that each agent's
\emph{expected} direction of movement at each time step is a descent direction,
even though it may not be a descent direction in actuality due to
stochasticity.}
and hence we can adapt proof by \cite{Tsitsiklis} to show that our model
converges to a local minimum under reasonable assumptions on how bacteria
communicate. Simulation studies indicate that the solution is feasible and
leads to improvements over prior methods and over single cell and single swarm
behavior. Finally, we discuss how these efficient and robust bacterial algorithms may be applicable to
distributed sensors or wireless networks that operate under strict
communication and computation constraints \cite{Canright}, \cite{Anastasi}.

\section*{Results}
\subsection*{A computational framework for understanding bacterial food search}

While bacterial food search has been extensively studied, at both molecular
and cellular levels, most studies have focused on the individual cell, rather
than characterizing the collective performance of a bacterial swarm. To formally analyze this process and to derive methods that can be used for
other tasks performed by severely restricted agents, we consider collective
bacterial food search as a distributed gradient descent (DGD) algorithm for
determining the direction of movement for each agent. DGD is a classic
distributed optimization algorithm for finding a minimum of an objective
function. In our case, the minimum corresponds to the food source, and the
objective function incorporates the terrain over which the bacteria cells search. DGD
is based on \emph{message passing} between nodes (individual cells or agents)
in a graph. These messages contain information that each node uses to update
its own movement direction and velocity. The goal of DGD is for all nodes to
converge to a single location in the search space.

While DGD has been studied in several different application areas,
there are several differences between standard DGD algorithms
\cite{Zinkevich,Li} and bacterial food search. First, classical algorithms
assume that message passing occurs on a fixed and static network, whereas for
bacteria edges and their weights are based on a sphere of influence which
changes over time based on each bacterium's current location. Second, classical
algorithms typically do not limit message complexity (in terms of message size
and number of messages sent), whereas bacteria have a limited message
vocabulary \cite{Cheong} and may attempt to minimize the number of messages
sent to conserve energy. Third, classical algorithms require
significant data aggregation at individual nodes, whereas bacteria are not
believed to collect and store data in this way. Considering that bacteria
have been solving this optimization problem under these constraints for billions of years, we hope to learn interesting algorithmic strategies
from studying this process. Such algorithms may address a new class of problems
that may be important in other applications, such as swarm
robotics~\cite{Rubenstein}.

For bacteria, the movement of an agent $i$ at time $n$ is a function $d_i(n)$
of two quantities: its own sense of direction $\theta_i(n)$ (based on the
chemical gradient), and the locations and movement directions of other cells in
the swarm. Nodes in this graph correspond to bacteria, at some current location
$\vec{x_i}(n)$ and edges, representing physical distance, exist
between two bacteria that lie within a sphere of influence of each other. The
challenge lies in updating $d_i(n)$ based on the individual belief and the beliefs
of neighbors, while using simple messaging (formalized below). For bacteria,
messages passed along edges in the graph contains both homophilic components
(attraction and orientation) and a heterophilic component (repulsion). Below,
we first present prior work by Shklarsh et al.~\cite{shklarsh} that describes
bacteria food search using a differential equation (DE) model and then present
our DGD model, which improves upon the Shklarsh DE model in performance while
relying on weaker, more biologically realistic, assumptions.


\subsection*{The Shklarsh model: Cell based differential equation computation}
Initial models for bacterial swarm movements assume that cells solve a system
of differential equations (DE) to determine their next move \cite{shklarsh}. We
briefly review this model below. The model assumes that individual cells follow
a chemical gradient of food source by decreasing their (random) tumbling
frequency at high concentrations and thus largely move in the
direction of the food. Specifically, the bacteria perpetually moves in a direction which it
repeatedly perturbs randomly. The frequency and magnitude of these
perturbations are inversely related to the change in the food concentration
between iterations, with the rough effect that the agent continues to move in
directions along the gradient. Formally, under these assumptions, at time $n$,
the bacterium changes its direction by an angle $\theta (n)$, which is a
function of $\Delta c(n)$, the difference in food concentration between the
current and previous time steps. Specifically, the new tumbling angle
$\theta(n)$ is sampled randomly from a Gaussian distribution
$\theta(n) \sim N(\theta(n - \Delta n), \sigma(\Delta c(n))^2)$ centered at the
previous angle $\theta(n - \Delta n)$, with the variance
$\sigma (\Delta c(n))^2$ given as:
\begin{equation}
\sigma (\Delta c(n))^2=\left\{
\begin{array}{ll}
0; & \Delta c(n) \geq 0\\
\pi; & \Delta c(n) < 0
\end{array}
\right..
\end{equation}
Thus, based only on its own perception of the food gradient, the $i^{th}$ agent
updates its location $\vec{x_i}(n)$ according to
\begin{equation}
\vec{x_i}(n+\Delta n)=\vec{x_i}(n)+\vec{v_i}(n) \cdot \Delta n,
\label{eq:single_cell_update}
\end{equation}
where $\vec{v_i}(n)$ is the unit vector in the direction of the movement.
\begin{equation}
\vec{v_i}(n)=(\cos(\theta (n)),\sin(\theta (n))).
\end{equation}

So far we have discussed movements based on sensing by individual cells
(agents). However, in almost all cases, cells move in large swarms.
Communication among swarm members, and between swarms, improves the ability of
individual cells to handle obstacles in the direction of the food source leading to faster and more efficient ways to reach the food. The communication
among agents is divided into three components: (1) {\em repulsion} from very
close  agents to avoid collision ; (2) {\em orientation} to match the
direction of neighboring cells, and (3) {\em attraction} to distant agents
to keep the swarm unified (Figure~\ref{model}(b)). The model of Shklarsh et al.
assumes that cells align their trajectory with the direction of the other cells
if they are close enough, while being attracted toward cells that are
relatively far away.

Denote by $\vec{u_i}(n)$ the vector agent $i$ computes
using the messages from the other cells (in this model, movements are of fixed
length and so the only variable in each iteration is the direction). Then, if
any other agents $j$ are within the physical interaction (repulsion) range $RR$
as shown in Figure~\ref{model}(b), the Shklarsh model sets:
\begin{equation}
\vec{u_i}(n)=-\sum_{j\in B_{RR}}\frac{\vec{x_i}(n)-\vec{x_j}(n)}{\|\vec{x_i}(n)-\vec{x_j}(n)\|}.
\end{equation}
Otherwise, the Shklarsh model sets:
\begin{equation}
\vec{u_i}(n)=\sum_{j \in B_{RO}}\vec{v}_j(n)+\sum_{j\in B_{RA},j\notin B_{RO}}
\frac{\vec{x_i}(n)-\vec{x_j}(n)}{\|\vec{x_i}(n)-\vec{x_j}(n)\|},
\label{eq:orient_plus_attract}
\end{equation}
where the first sum is over all agents $j$ in the orientation range $RO$ and
the second sum is over all agents in the attraction range $RA$ (but not in the
range $RO$). Next, the agent combines the messages it received with its own
observation, resulting in the following modification of equation
(\ref{eq:single_cell_update}):
\begin{equation}
\vec{x_i}(n+\Delta n)
  = \vec{x_i}(n)
    + \left( \frac{\vec{u_i}(n)}{\|\vec{u_i}(n)\|}+w\vec{v_i}(n) \right) \cdot \Delta n,
\end{equation}
where $w$ is a scalar weighting factor.

\subsection*{A distributed gradient descent model}
While the model presented by Shklarsh et al. \cite{shklarsh}
captures the basics of bacterial swarm movements, it suffers from
several problems which make it unlikely to be used by real cells and, as we show in
Results, less efficient. First, the model assumes that cells can
determine their exact distances from \emph{each} other cell (as can be seen by
the summations over neighbors cells in the orientation and attraction ranges).
This implicitly assumes that cells can {\em identify} individual senders, which
is very unlikely given the large and dynamic nature of bacterial swarms. In
addition, the model assumes that cells can interpret complex (real-valued)
messages regarding the locations and orientation of other cells, which is also
unrealistic \cite{Cheong}. Finally, the model assumes that, within each of the
ranges above, each cell exerts the same influence regardless of their distance
from the receiving cell, which is again unrealistic due to the nature of the
communication channel (diffusion of a secreted protein). We have thus modeled
bacterial food search using a DGD model that relaxes many of these assumptions
while still allowing cells to (probably) reach an agreement regarding the
direction of movement and eventually the location of the food source, as
observed in nature.

Our model still distinguishes between physical interactions (leading to
repulsion) and messages (secreted proteins). The former is handled by summing
up the number of cells that are in physical proximity without relying on their
exact location. However, we make several changes to Equation
(\ref{eq:orient_plus_attract}). First, we remove the requirement that cells
identify the distance and direction to each other cell (and thus determine
whether to use the attraction or orientation terms). Instead, we simply sum
over all cells, taking into account their relative influences under the
assumption that message strength decays exponentially with distance
\cite{Wartlick}. Second, we discretize the messages that cells receive,
resulting in simple messages with finitely many possible values. The changes
lead to the following modification of equation (\ref{eq:orient_plus_attract}):
\begin{small}
\begin{equation}
\vec{u_i}(n)
  = D_{L,T}\left(\sum_j \exp(-(C_o \|\vec{x}_i(n)-\vec{x}_j(n))\|) \vec{v}_j(n)
                 + \sum_j
\frac{\exp(-(C_a \|\vec{x}_i(n)-\vec{x}_j(n))\|)(\vec{x}_i(n)-\vec{x}_j(n))}
     {\|\vec{x}_i(n)-\vec{x}_j(n)\|} \right).
\end{equation}
\end{small}
Here, $D_{L,T}$ is a discrete thresholding operator parametrized by $L$, a
positive integer denoting the number of possible messages, and $T$, an upper
bound above which all messages as treated as the highest value
possible (see \cite{Emek} for the exact construction of this ``stone-age
computing'' threshold which has been used in ant models). $C_a$ and $C_o$ are
positive diffusion constants, determining how quickly the attraction and
orientation signals diffuse from the source agent. Typically, $C_o > C_a$, in
which case nearby agents are influenced more by the orientation component and
far away agents are influenced more by the attraction component.
Under this model, bacteria communicate orientation and attraction information
using only $3 + \log_2 L$ bits to communicate ($\log_2 8$ bits for direction
and $\log_2 L$ bits for magnitude). In addition, we also add a Gaussian noise
component with a small variance $\sigma$ to make this process stochastic. Note
that the individual component of the agents' movement (based on the chemical
gradient they perceive) is identical to that of the Shklarsh model, and that we
are modifying only the communication model. See Supplement for the complete
combined model.

\subsection*{Convergence theorem}
Unlike standard distributed gradient descent algorithms, the model described
above does not rely on a fixed network. Instead, in each iteration, the
topology of the network (and thus the weights placed on neighbors)
changes with their movement. To prove that using such network (and the
computation we assume) indeed leads to convergence of the swarm as seen in
nature (regardless of agents' starting locations), we adapt a convergence
theorem for distributed pseudogradient descent from \cite{Tsitsiklis}. The
convergence proof is only focused on the attraction component of the
model (see Discussion). However, the theorem holds for both synchronous and
asynchronous settings, and thus our model could be generalized to the
case where agents' messages themselves travel stochastically.

The convergence theorem is fully stated and proven in the Supplement. The main idea of the proof
is that there exists a sequence $\{y(n)\}_{n = 1}^\infty$ such that, if, at
time $n$, all agents were to cease to follow the gradient while continuing to
move according to the attraction term (i.e., weighted averaging), then all
agents would converge to location $y(n)$. For example, in the case that all agents
communicate with equal weights, $y(n)$ is simply the mean of the agents'
positions at time $n$. Under reasonable assumptions on the edge weights,
information from each agent is likely to propagate throughout the
swarm and so the agents will converge on $y(n)$ (plus individual
noise), regardless of their starting position. Furthermore, once the
agents are sufficiently close, the change in $y(n)$ (which is a
weighted average of the gradients perceived by each agent) in each
iteration becomes approximately proportional to $J(y(n))$, so that
the swarm collectively behaves as a traditional gradient descent. 

The proof relies on two lemmas stated in the Supplement. The first 
claims that agents continue move in the
correct descent direction once they have detected an increase in
food gradient in one of their tumbles (until that direction ceases
to be a descent direction). As we show, to achieve such expected
decrease as required by the lemma for each step, we need to change
either the tumbling distribution assumed in the original model, or the
number of tumbles per step (i.e. prior to communicating a new
location). While the former (a uniform tumbling distribution) is less likely
in practice, the latter (communicating only when a new descent direction is
established) both makes biological sense \cite{Hense} and, as we show
empirically in Results, reduces the time it takes the
swarm to reach the food source. For the second lemma we again
need to modify the original algorithm so that the step size (amount
of progress made in the direction computed) is proportional to the detected
gradient, causing the agents to slow as they approach the food source. This may
be implemented in practice via a feedback loop used by bacterial cells
\cite{Vladimirov}. Detailed assumptions and rigorous proofs of both lemmas and
the convergence theorem are given in the Supplement.

\subsection*{Empirical Results}

To determine whether the restricted communication model we assume
can indeed lead to efficient convergence we performed
several simulations of bacterial food search. First, we compare the
search efficiency of bacterial food search with and without
communication, and between our model and the Shklarsh et al. model.
Second, we introduce multiple swarms and test how their trajectories
affect each other. Third, we explore the predictions of
the model for a setting in which a fraction of the bacteria are
behaving differently than the others. This latter point is of great current
interest since `cheaters' (cells that receive messages
but do not spend the energy on sending them) may be responsible for
a form of antibiotic resistance that has been recently observed
\cite{Yurtsev}. Finally, we evaluate our method
using new experimental data.  Thus, the results presented below are of interest to
both the computational part of this work (efficient and robust DGD
model) and the biological aspects (models of bacterial
coordination).

To save space, in the remainder of this section we refer to the Shklarsh et al.
model as the `DE' model and our model as the `DGD' model (even though both
rely on updating magnitude and direction in time steps).

\subsection*{Performance on a realistic food search simulation}
To evaluate the performance of our method we first tested it using
the terrain and obstacle model from Shklarsh et al. \cite{shklarsh}
(see Supplement). In these settings we varied the number of agents,
the communication between agents and the number of swarms.

The quantity we compared was the time
it took cells to reach the food source (in terms of steps, since
both algorithms can be run synchronously).
Figure \ref{single_swarm} presents the distribution of the number of
steps it takes cells to reach the food source under several different models.

The first is a model without communication (i.e. each cell can only sense the
food gradient, but does not receive or send any message). The second is the
Shklarsh et al. model with adaptive weighting, and the other two models are
our DGD model with fixed or variable step size (the latter is required for the convergence proof above  while the former is usually used in bacterial models).
As can be seen, communication greatly improves the time it takes cells to
reach the food source, which may explain why such a secretion-based
communication system has evolved in this species. As for the specific
communication model, the DGD model improves the results when compared to the DE
model, even though our DGD model severely restricts the set of messages that
can be used. In fact, the discretization of messages decreases both the mean and
variance of the distribution. This is likely due to the fact that, by
thresholding, the discretization step is effectively reducing the large noise that can be associated with individual messages. In addition, the
distance-weighted edges (corresponding to the diffusion rates of secreted
communication molecules) also improve the performance of the method. Simulation movies with and without communication between cells can be observed on the Supporting Website.

Figure \ref{seq} displays the effect on the time it takes cells to reach the
food source if another swarm is added to the simulation. In this sequential set
up, the second swarm starts 50 iterations after the first swarm. 
As can be seen, the fact that the first swarm was already
able to successfully navigate to the food source enables the second swarm to
utilize (at least partially) the trajectory they identified to further reduce
the time it takes to reach the food. Interestingly, while the improvement for
the second swarm is indeed large, we also see a {\em decrease} in the
performance of the first swarm compared to the single swarm result presented in
Figure \ref{single_swarm}. This is due to the
negative influence that the
second swarm has on the first when it enters the region of influence. Since the
second swarm starts in the opposite direction of the food source, the first
swarm is (partially) adjusting its direction incorrectly (based on
attraction to cells in the second swarm) increasing the number of
iterations it takes cells to reach the food source.

\subsection*{Sensitivity to silent agents}
Previous work has shown that some cells in a population become
`silent' \cite{velicer},\cite{rainey}, \cite{strassmann}. These
cells receive messages from the other cells but do not send messages
themselves. While such behavior is beneficial from the individual
standpoint (less energy is required to synthesize and secrete the signaling molecules) it may be
harmful for the population as a whole since if these `silent' cells
proliferate the population will lose its ability to utilize
communication to improve food search.  Recent work has shown that
stochastic activation of such silencing and other individual
behavior mechanisms can explain how they can be advantageously used
(for example, for developing antibiotics resistance) without affecting the
overall ability of cellular coordination. We have
thus used the obstacle model again to study the  sensitivity of
swarm performance to the fraction of silent cells in the population.
For this, we varied the fraction of silent agents from 0 to 1. As
can be seen in Figure \ref{silent} up to a certain threshold we do
not see a large impact for the increase in silent cells which
supports the recent findings of \cite{Yurtsev}. Specifically, based
on our models the performance of the population is only $20\%$ less than optimal
even if $85\%$ of the cells are silent. 

We have further analyzed various aspects of the communication model to
determine the roles of each type of message being sent (orientation,
attraction) or sensed (physical proximity), in terms of the impact on the time
taken to reach the food source. See Supplement for details.

\subsection*{New experimental data: Chemotactic migration of \textit{E. coli} cells with or without cell-cell communication}
We designed and fabricated a microfluidic device to test aspects of our model. We studied the chemotactic migration behavior of wild type (wt) and $\Delta tsr$ mutant \textit{E. coli} cells in the microchambers in the presence of chemoattractant gradient. While the initial modeling assumed a swarm of agents at the start, due to the technical constraints, in our experiments the cells were evenly distributed in the microchambers at the beginning. After the media with chemoattractant was introduced in the main channel, gradients of the chemoattractant were generated in the microchambers, with high concentration of the chemoattractant at the inlet corner and lower concentration at the other three corners (Figure ~\ref{Experiment}). As seen in the supplementary movies and Figure \ref{Experiment}b), the \textit{E. coli} cells in the microchambers sense the gradient and migrate toward the inlet corner before escaping into the main channel. 
Previous studies \cite{park} confirmed that motility buffer and the deletion of the Tsr chemoreceptor would not decrease the motility and chemotactic migration speed of \textit{E. coli} cells to Aspartic acid (which engages the Tar chemoreceptor). However, cells in the nutrient-deprived motility buffer cannot secrete the amino acid glycine while cells lacking the Tsr receptor cannot sense the gradient of glycine. It has been previously shown that glycine plays a key role for several collective behaviors of starved \textit{E. coli} cells. In a nutrient-depleted environment, the \textit{E. coli} RP437 cells can sense and be attracted by glycine secreted by other cells forming dense aggregates \cite{park}. 
Indeed, in our experiments, WT \textit{E. coli} cells in M9CG medium migrate much faster than wt cells in motility buffer and $\Delta tsr$ mutant cells in M9CG. More than $90\%$ of wt cells escaped from the microchamber in 10 minutes in M9CG, while only $60\%$ cells escaped after 60 minute exposure to the chemoattractant gradient in the two control experiments (Figure S6). 
As the cells in the two control experiments can neither secrete glycine (in motility buffer) nor sense glycine ($\Delta tsr$ cells) no cell-cell communication is expected and observed in these experiments. This result agrees with our model assumptions about the importance of cell to cell communication in bacterial food search. 

\subsection*{Comparing model based simulations with experimental data}
To further test the DGD model and compare it with the Shklarash DE model, we used both models to simulate bacterial food search with the same initial configuration as the experimental setting (initially cells are evenly distributed rather than grouped in a swarm). We next performed analysis to determine the ratio of the number of cells at the food source (bottom right corner) to the number of cells at the opposite corner for each minute / iteration in the experiment and simulations, respectively (Methods). This process was repeated for the three different obstacle sets we tested (these varied in their sizes for each experiment but the overall coverage remained the same for all obstacle sizes (Methods). The results, comparing the two models and the experiments are presented in Figure ~\ref{Experiment} (each simulation curve is averaged over 50 random starts). As can be seen, and as expected, both models and experiments agreed that the ratio increases as time / iteration increase due to cell attraction to the food source. The models also correctly agreed on the ordering of the different obstacle sizes in terms of the ratio (recall that the coverage is the same, and so bigger obstacles also mean fewer overall obstacles allowing for easier navigation if communication is allowed). However, the models differed in the slope of the predicted ratio increase and in the similarity between the different obstacle sizes. Overall, we observe that the DGD model leads to slopes that are in better agreement with the experimental results. In addition,  in the DGD model the two larger obstacle setups are grouped together and are separated from the smallest whereas in the DE model the two smaller settings are grouped together and separated from the largest. Here again the experimental results correlate more closely with the DGD model.

\section*{Discussion and conclusions}
We have shown, both theoretically and in simulations and experiments, that a distributed
gradient descent model can efficiently interpret the communication of agents under severe limitations. These include limits on the complexity of messages and the ability to identify which agent is sending the message, while at the same time
assuming a dynamic environment where neighbors' locations (and their influence)
change constantly. We have proved that, under reasonable biological assumptions,
the communication algorithm discussed is likely to converge, helping to explain how bacteria can efficiently coordinate food search in harsh
environments  and and improving upon prior models of this process. Moreover, the social dynamics of bacteria can also affect their ability to resist antimicrobial therapy, and inhibition of bacterial cooperation is an alternative approach to minimize collective resistance \cite{vega}.

Our convergence proof only holds for the attraction information and does not hold for the repulsion and direction components of the computation performed by
each agent. Since the orientation information is also a vector averaged over all neighbors, we believe that the proof can be extended to include this
communication term as well, though a key challenge would be to understand the
sometimes competing goals of local versus global improvement near certain
obstacles. Repulsion is more difficult to analyze because it is not based on
averages. However, as we show in the Supplement, the attraction term has by far
the most significant effect on swarm performance while the other two
communication terms have a much less significant impact on the time it takes cells to reach
the food source. Hence, for computational applications of our method, it may
suffice to use the attraction term, in which case convergence is guaranteed.

\newpage
\section*{Methods}

\subsection*{Bacterial strains and growth conditions}
The wild-type \textit{E. coli} RP437 strain and its derivative HCB317 ($\Delta tsr$, from the lab of Prof. Howard Berg), were used throughout the study. Cultures were launched from a frozen stock and grown overnight at $30^{\circ} C$ with agitation at $240rpm$ in M9 minimal medium supplemented with $1g/l$ casamino acids and $4g/l$ glucose (M9CG). We then diluted the overnight culture hundredfold in the morning with fresh M9CGand the cells were harvested at early log phase ($OD600nm = 0.1$).  Prior to experiments we  centrifuged, washed and resuspended the cells in M9CG or motility buffer (10 mM potassium phosphate, 10 mM sodium lactate, 0.1 mM EDTA, and 1 mM L-methionine, pH=7.0). We added chloramphenicol  ($30\mu g/ml$) to all media to maintain the plasmid expressing yellow fluorescent protein (YFP) or red fluorescent protein (tdTomato) constitutively in the cells.
\subsection*{Preparation of microfluidic device}
We designed a microfluidic device that contain three different designs of evenly distributed micropillars (obstacles) within $1000\times 1000 \mu m^2$ microchambers that in their totality occupy $25\%$ of the total surface area (Figure ~\ref{Experiment} a)). All chambers are connected to a wide channel through a $5\mu m$ wide channel to allow for the introduction of cells and media. The layout of the device is shown in Figure ~\ref{Experiment}a).

We fabricated the microdevice using standard soft lithography. Briefly, $10\mu m$-thick photoresist SU-8 2010 (MicroChem, Newton, MA) was spin-coated onto a polished silicon wafer. The spin-coated wafer was then exposed to UV light through the photomask by Karl Suss MJB3 aligner. Unexposed resist was removed with a SU-8 developer so that a raised surface remained that was the negative of the desired structure. The microfluidic device was cast from the biologically inert polymer PDMS (Sylgard 184, Dow Corning), which was heat-cured on the mold at $65^{\circ} C$ and then peeled away. After inlet-outlet holes were cut, the PDMS chips were briefly treated in air plasma cleaner for 30 seconds to render them hydrophilic and to enable the PDMS to seal to a thin glass coverslip. Immediately after the plasma treatment and bonding, $5\mu L$ of $20mg/mL$ BSA solution was added into the two holes in the PDMS. Both holes were then sealed with masking tape. The device was kept at room temperature for at least one hour to coat the PDMS channel walls with a thin layer of BSA. The BSA solution in the microchannels and microchambers was replaced with M9CG or motility buffer after surface passivation.

The motile cells which were resuspended in M9CG or motility buffer were then introduced into the main channel and was allowed to swim continuously into the microchambers. After reaching the desired cell density in the microchambers, fresh M9CG or motility buffer containing $300\mu M$ L-aspartic acid (MP Biomedicals) as chemoattractant was pumped into the main channel at a flow rate of $5\mu L/min$.
\subsection*{Time-lapse imaging and data analysis}

The chemotactic response of the fluorescently labeled \textit{E. coli} cells was observed and recorded on a fully automated inverted microscope (Zeiss AxioObserver Z1) equipped with a motorized stage (Applied Scientific Instruments) and an external LED light source (Metaphaser MP-LE1007) for fluorescence. The time-lapse movies were collected every minute with a CCD camera (Zeiss AxioCam MRm) at room temperature ($\sim 27^{\circ}C$). The movies were processed for display with ImageJ software. The cell number in the microchambers at each image frame was counted automatically by a custom pipeline with the open-source software Cellprofiler. 

\subsection*{Analysis of movies to identify concentration at food source}
\addtocounter{subsection}{-1}

\begin{figure}
\centering

\includegraphics[scale=.6,trim=0 4 0 0,clip]{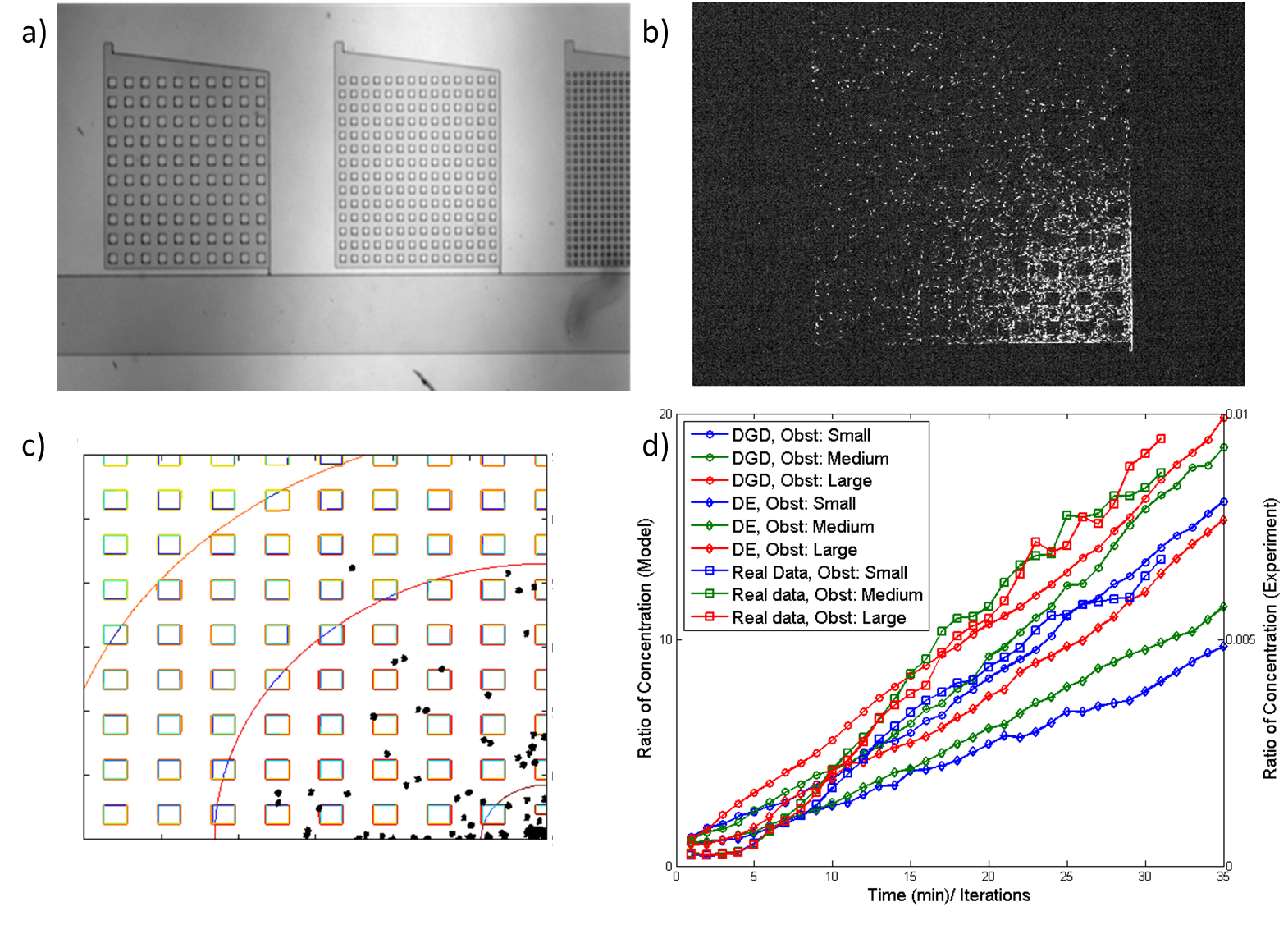}
\caption {Experimental validation of  the model a) Experimental setup. Microchambers with three different obstacle patterns though with the same overall coverage, b) Snapshot of chemotaxis movie, showing higher concentration of cells at the food source inlet (bottom right). Obstacles can be observed as dark rectangles surrounded by white cells. , c) Simulation setup mimicking experiment, d) Comparison of DGD and DE models with experimental data. As can be seen, the DGD simulation results lead to slopes and groupings  that are in better agreement with the experimental results.  }

\label{Experiment}
\end{figure}

We quantified the time it takes cells to reach the food source by tracking the ratio of the cell concentration at the bottom right corner (chemoattractant inlet) and top left corner (lowest concentration of chemoattractant) of the chamber Figures ~\ref{Experiment}a)and b). For each corner we considered a quarter circular area of fixed radius (200 pixels) to measure cell concentration. In the biological experiments, after reaching the food source cells escape through the inlet. Therefore the total number of cells in the chamber decreased with time. To allow for comparison with the simulation results in which the total number of cells in the chamber is fixed we normalized the ratio of cell concentrations in the experiments by total cell count at each time point.

\end{doublespace}

\begin{figure}
\centering

\includegraphics[scale=.7]{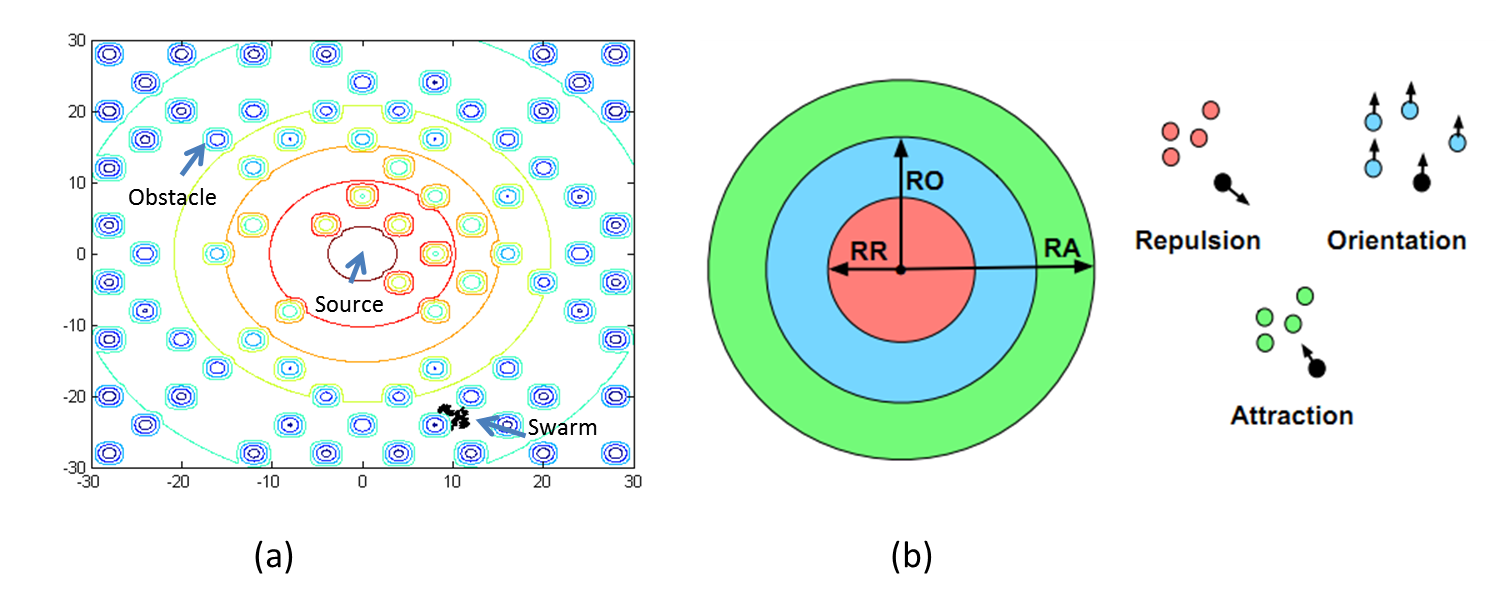}
\caption{{\bf (a):} Terrain model for bacterial food search.
Obstacles are randomly placed and the food source is at the center
of the region. Contours display the diffusion of the food source
gradient. {\bf (b):} Dynamics of repulsion, orientation, and
attraction for a single bacterial cell in the Shklarsh et al. model.
RR = radius of repulsion, RO = radius of orientation, and RA = radius
of attraction. While we still maintain the physical (RR) versus
communication (RO, RA) split between the repulsion and attraction /
orientation information, our model does not assume that the identity
of the sender is known and so it  does not distinguish cells in the
RO and RA locations.}

\label{model}
\end{figure}

\begin{figure}
\centering
\includegraphics[scale=1,trim=0 8 0 0,clip]{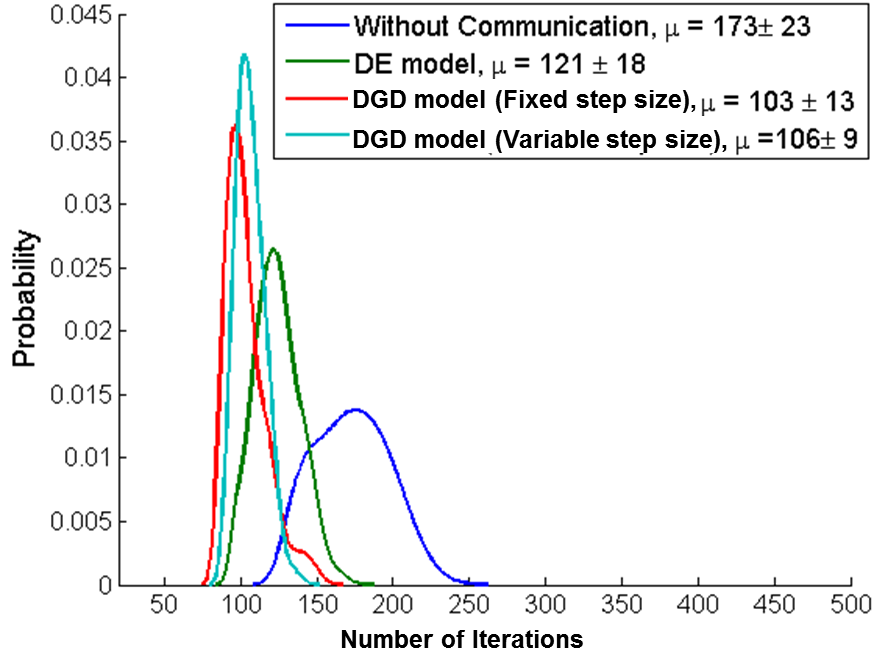}
\caption{Comparison between the original Shklarsh model and the DGD model.
$\mu$ denotes the mean number of iterations per agent, plus or minus the
standard deviation.}\vspace{-15pt}\label{single_swarm}
\end{figure}

\begin{figure}
\centering

\includegraphics[scale=1,trim=0 8 0 0,clip]{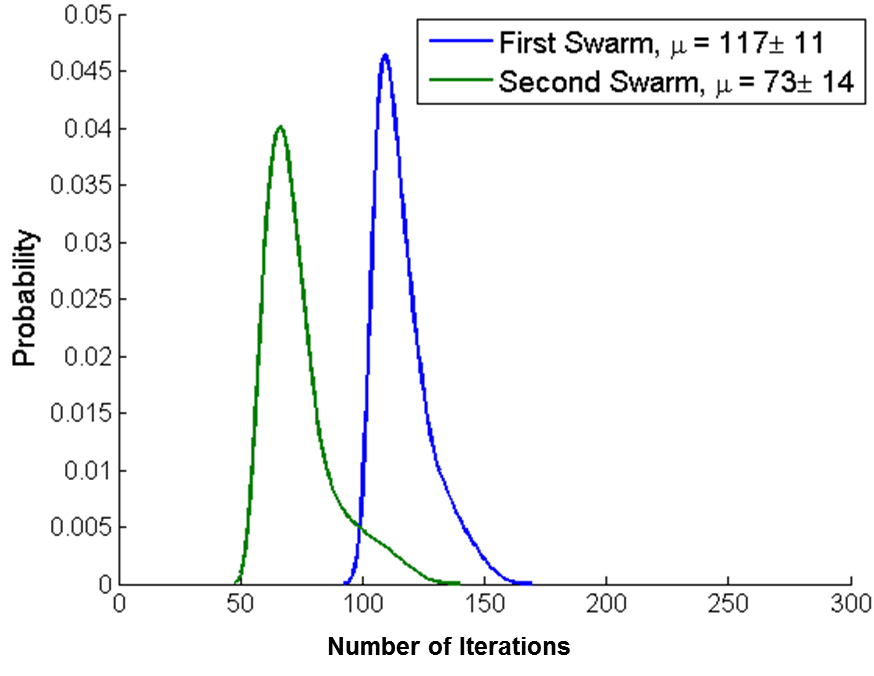}
\caption{Distribution of number of iterations for two sequential swarms using
variable step sizes. $\mu$ denotes the mean number of iterations per agent.}
\label{seq}
\end{figure}

\begin{figure}
\centering
\includegraphics[scale=1,trim=0 4 0 0,clip]{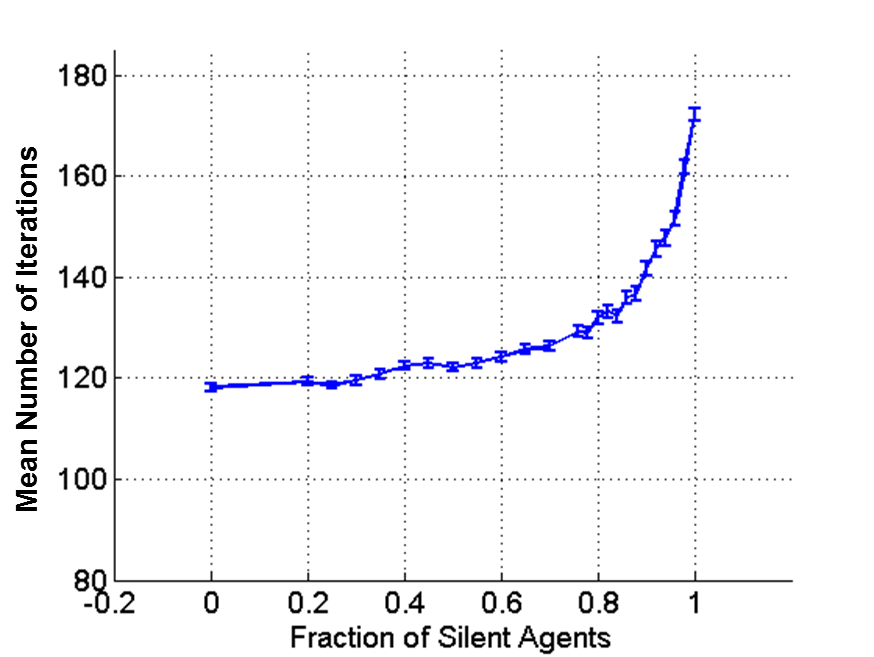}
\caption{Mean path lengths over silent fractions, based on 100 trials. More
silent bins were taken in the range of .7 to 1 to illustrate the phase
transition. Error bars indicate standard deviations.}
\label{silent}
\end{figure}

\newpage
\textbf{Figure ~\ref{model} Environment and model assumptions}
\newpage
\textbf{Figure ~\ref{single_swarm}: Comparison of search time under different communication models}
\newpage
\textbf{Figure ~\ref{seq}: Performance of multiple swarms}
\newpage
\textbf{Figure ~\ref{silent}: The effects of silent agents}
\newpage
\textbf{Figure ~\ref{Experiment}: Experimental results}
\end{document}


\pagestyle{plain}

\mainmatter
\title{Supplementary Information: Distributed Gradient Descent in Bacterial Food Search}

\author{Shashank Singh$^{*1}$, Sabrina Rashid$^{*2}$, Zhicheng Long$^{3}$, Saket Navlakha$^4$, Hanna Salman$^5$, Zolt\'{a}n N. Oltvai$^6$, Ziv Bar-Joseph$^{7\dagger}$}
\institute{$^1$ Machine Learning Department and Department of Statistics, Carnegie Mellon University, Pittsburgh, PA 15213.\\
$^2$ Computational Biology Department, Carnegie Mellon University, Pittsburgh, PA 15213.\\
$^3$ Department of Pathology, Department of Physics and Astronomy, University of Pittsburgh, Pittsburgh, PA 15213.\\
$^4$ The Salk Institute for Biological Studies, Center for Integrative Biology, La Jolla, CA 92037.\\
$^5$Department of Physics and Astronomy, Department of Computational and Systems Biology, University of Pittsburgh, Pittsburgh, PA 15260.\\
$^6$Department of Pathology, Department of Computational and Systems Biology, University of Pittsburgh, Pittsburgh, PA 15260.\\
$^7$ Machine Learning Department and Computational Biology
Department, Carnegie Mellon University, Pittsburgh, PA 15213.\\
$^*$ These authors contributed equally.\\
$^\dagger$ Contact author. Email: zivbj@cs.cmu.edu}
\maketitle
\section*{ Supporting Methods}
\subsection*{Complete update rule for a single cell}
Below we present the complete update rule performed by each cell in
each of the steps of the algorithm to compute its new trajectory. At
each iteration, there are three communication components in the BP
model, i) repulsion, ii) orientation, and iii) attraction.

As discussed in the main text, we divide the update into two cases. If
there are {\em any} cells within the Repulsion Radius (RR,
determined by physical interactions) then the cell moves as follows:
\begin{equation}
\vec{u_i}(n)
  = -\sum_{j \in B_{RR}}
              \frac{\vec{x_i}(n)-\vec{x_j}(n)}{\|\vec{x_i}(n)-\vec{x_j}(n)\|},
\end{equation}
were $x_i(n)$ is the location of agent $i$ at time $n$. Otherwise (if there are
no cells in RR) we set:
\begin{small}
\begin{equation}
\vec{u_i}(n)
  = D_{L,T}\left(\sum_j \exp(-(C_o \|\vec{x}_i(n)-\vec{x}_j(n))\|) \vec{v}_j(n)
                 + \sum_j
\frac{\exp(-(C_a \|\vec{x}_i(n)-\vec{x}_j(n))\|)(\vec{x}_i(n)-\vec{x}_j(n))}
     {\|\vec{x}_i(n)-\vec{x}_j(n)\|} \right).
\end{equation}
\end{small}
In either case, the cell updates its own position by moving according to
\[x_i(n+1)=x_i(n)+\gamma_i(n)(u_i(n)+s^i(n)),\]
where $s^i(n)$ is the belief of cell $i$ at time $n$ based on its sensing of
the food gradient ($s^i(n)$ is a pseudogradient, as shown in Lemma 1 below).
$\gamma_i(n)$ is the step size for agent $i$ at time $n$, which, as discussed
in the main text, is a function of the food gradient at the cell's location
(a detailed formulation of $\gamma_i(n)$ is in the proof of Lemma 2 below).

\subsection*{Convergence Theorem}
In a simplified model that incorporates only the attraction term of
communication, our convergence theorem follows from a general convergence
theorem proven by Tsitsiklis et al. for a broad class of distributed
pseudogradient descent algorithms (Theorem 3.1 and Corollary 3.1 of
\cite{Tsitsiklis}). Work is needed only to show that our setting satisfies the
six assumptions therein (listed and justified below). In order to match the
notation of \cite{Tsitsiklis}, we note that the update equation (using only the
attraction term) can be written in the form
\begin{equation}
x_i(n+1)=\sum_j^M a_{ij}(n)x_j(n)+\gamma_i(n)s^i(n)
\label{eq:true_update_app}
\end{equation}
where
$a_{ij}(n)$ is the edge weight between cells $i$ and $j$ at time $n$.

{\bf Convergence Theorem:}
Let $x_i(n) \in \R^2$ denote the position of the $i^{th}$ agent at time $n$ and
let $J : \R^2 \to \R$ denotes the concentration of the food source (the
objective function). Under assumptions A1-A6 (see below), the following
hold:
\begin{enumerate}
\item  There exists a constant $\gamma^* > 0$ such that, if each step size is
at most $\gamma^*$ (i.e., $\sup_{i,n} \gamma_i(n) \leq \gamma^*$), then,
with probability $1$, for all distinct agents $i$ and $j$,
\[\lim_{n \to \infty} (x_i(n) - x_j(n)) = 0.\]
\item If, in addition, the level sets of $J$ are compact, then, with
probability $1$, for all agents $i$,
\[\lim_{n \to \infty} \|\nabla_x J(x_i(n))\|_2 = 0.\]
\item Finally, if, in addition, all stationary points of $J$ are minima (e.g.,
if $J$ is convex), then, with probability $1$, for all agents $i$,
\[\lim_{n \to \infty} J(x_i(n)) = \inf_x J(x).\]
\end{enumerate}

The three components of the above theorem gives sufficient conditions for the
agents to converge spatially, to converge to a connected set of  stationary
points, and to converge to a connected set of minima, respectively.
Assumptions A1-A4 (stated precisely below) are that $J$ is
sufficiently smooth (its gradient is Lipschitz continuous) and that the agents'
communication network is sufficiently dense so that information from any agent
eventually propagates through the swarm. Assumptions A5 and A6 correspond to
Lemmas 1 and 2 which we explicitly discuss below.

The lemmas, stated below, are needed to establish that our algorithm is indeed a distributed
pseudogradient algorithm before we can apply the results of \cite{Tsitsiklis}.
The lemmas essentially state that (a) the agents expect to move in a descent
direction and (b) the variance in the agent's movements is at most proportional
to the magnitude of the true gradient (so that, for example, it vanishes near
stationary points and hence agents cease to explore once they have located an
optima).

\subsubsection*{Assumptions}
The assumptions needed for the general convergence proof are the following:

\begin{enumerate}
\item[\bf A1] There is some $\alpha > 0$ such that $a_{ij}(n) \geq \alpha$, for all
times $n$ and distinct agents $i$ and $j$.
\item[\bf A2] There exists a constant $B$ such that the time between
consecutive communications between any pair of distinct agents $i$ and $j$ is
at most $B_1$ \cite{Nancy},\cite{Francis}.
\item[\bf A3] The number of messages communications between any two distinct
agents during any duration of length $B_1$ is bounded by some constant $B_2$.
\item[\bf A4] $J$ is continuously differentiable and its gradient $\nabla_x J$ is
Lipschitz continuous.
\item[\bf A5] The conclusion of Lemma 1.
\item[\bf A6] The conclusion of Lemma 2.
\end{enumerate}

Below we discuss why each of these holds in our setting. \\

{\bf A1} Since the edge weight $a_{ij}(n)$ is a positive, strictly decreasing
function of the distance $\|x_i(n) - x_j(n)\|_2$ and the agents
move only within a bounded region of $\R^2$, such an
$\alpha = \inf_{i \neq j, n} a_{ij}(n) > 0$ exists with high probability.

{\bf A2, A3} This point is somewhat subtle, and quite important for the
generality of our work. In our model, each agent regularly (after a fixed
number of rounds on individual movement, as discussed in the proof of Lemma 1
below) measures the \emph{weighted average} of the signals from all other
agents in the swarm, through the update equation (\ref{eq:true_update_app}).
The key observation here is that, in \cite{Tsitsiklis} (which only considers
communication as being between pairs of agents, and hence phrases the
assumption as above), each agent's update depends
\emph{only on the weighted average} of the messages received. Thus, this
assumption is \emph{weaker than} the assumption that the weighted average in
the update equation incorporates \emph{all} other agents (with weights
satisfying the constraints of assumption A1). By construction, our model
satisfies this.
\footnote{Compare \emph{Example V} of \cite{Tsitsiklis}, which considers a
similar communication pattern in a more abstract setting.}

{\bf A4} Food is assumed to diffuse smoothly from its source (usually according
to a Bessel, Gaussian, or exponential decay function of distance).

{\bf Lemma 1} Conditioned on the history of the algorithm, the
expected gradient perceived directly by each agent (not accounting
for information from the swarm) is a descent direction. That is,
$$E \left[ \frac{dJ}{dx}(x_i(n))s^i(n) \right] \leq 0.$$
In order to guarantee this we will allow agents to tumble a fixed number of
times between rounds of communication, so that, in expectation, they will find
and move along a descent direction.

\emph{Proof of Lemma 1:} We first show the case when the direction after a
tumble is chosen uniformly at random, in which case, the proof is
straightforward. For sake of generality, we then show the case where the new
angle is chosen according to any distribution (such a Gaussian) satisfying a
certain weak uniformity condition, allowing multiple tumbles within each round
of communication.

\emph{Uniform Case:}
In each iteration $n > 1$, if the previous direction $s^i(n - 1)$ is not a
descent direction (i.e., if $J(x_i(n)) \leq J(x_i(n - 1)$)), then
$\frac{s^i(n)}{\|s^i(n)\|_2} \sim \operatorname{Unif}(\partial B_1(0))$,
\footnote{$B_1(0)$ denotes the unit ball centered at $0$, and $\partial B_1(0)$
denotes it boundary.}
and the speed $\|s^i(n)\|_2$ is deterministic given $s^i(n - 1)$ and
$x_i(n - 1)$. Since the function
$s \mapsto \nabla_x J(x_i(n)) \cdot s$ is linear, its expectation over a
uniformly distributed variable is $0$, and so
\[\E\left[ \nabla_x J(x_i(n)) \cdot s^i(n) \middle| F_n \right]
    = \|s^i(n)\|_2 \E\left[ \nabla_x J(x_i(n))
                        \cdot \frac{s^i(n)}{\|s^i(n)\|_2} \middle| F_n \right]
    = 0.\]
If the previous direction is a descent direction, the agent retains its
previous direction. \qed

\emph{General Case:}
Now consider a more general algorithm parameterized by a \emph{tumbling
distribution} with density $D$ on $[-\pi,\pi]$ (above, $D$ is uniform). In
general, a single tumble may be insufficient to ensure that the expected
resulting direction is a descent direction, and so multiple tumbles may be
necessary. We assume that $D$ is somewhat uniform in the following sense:
\[0 < c := \inf_{\theta \in [-\pi,\pi)}
                    \int_\theta^{\theta + \pi/2} D(\phi_{2\pi}) \, d\phi,\;
\footnote {For $x \in \R$, $x_\pi = ((x + \pi) \mod 2\pi) - \pi$ denotes the
angle in $[-\pi,\pi]$ equivalent to $x$.}\]
i.e., the agent has probability at least $c \in (0,1/4]$ of tumbling toward
any particular quadrant. This assumption certainly holds for the Gaussian
tumbling distribution we use in the main paper, but also allows a broad range
of other possibilities, including any distribution whose density is lower
bounded away from $0$ on $[-\pi, \pi]$.

To prove the lemma, let $\theta_0 = \theta(\nabla_x J(x_i(n)), s^i)$,
$\theta_{\ell + 1} = \theta_\ell + \Delta \theta_\ell$, where each
$\Delta \theta_\ell \sim D$ denotes the change in angle due to the $\ell^{th}$
tumble.
\footnote{For two vectors $u$ and $v$,
$\theta(u,v) = \cos\inv\left( \frac{u \cdot v}{\|u\|_2\|v\|_2} \right)$ denotes
the (smallest) angle between $u$ and $v$. For notational convenience, we
measure angles as lying in $[-\pi,\pi]$, with $0$ denoting the direction of
$\nabla_x J(x_i(n))$.}
With probability in $(c, 1 - 3c)$, any particular
$|\theta_\ell| \leq \frac{\pi}{4}$, in which case
\[\nabla_x J(x_i(n)) \cdot s^i(n)
    \geq \frac{1}{\sqrt2} \|J(x_i(n))\|_2\|s^i(n)\|_2.\]
Also, with probability in $(2c, 1 - 2c)$,
$|\theta_\ell| \geq \frac{\pi}{2}$ for all $\ell \in \{1,\dots,k\}$, in which
case, by the Cauchy-Schwarz inequality,
$\nabla_x J(x_i(n)) \cdot s^i(n) \geq - \|J(x_i(n))\|_2\|s^i(n)\|_2$.
Otherwise, $|\theta_\ell| \in \left( \frac{\pi}{4},\frac\pi2 \right)$, so
$\nabla_x J(x_i(n)) \cdot s^i(n) \geq 0$.

Let $L \in \{0,\dots,k\}$ denote the last $\ell$ at which
$\theta_\ell > \frac{\pi}{2}$ (i.e., after which, the agent maintains its
current direction for the remaining $k - \ell$ tumbles). Since $L = \ell$ if
and only if $|\theta_0,\dots,\theta_\ell| \geq \frac{\pi}{2}$
$\theta_\ell$
\begin{align*}
\E\left[ \nabla_x J(x_i(n)) \cdot s^i(n) \middle| F_n \right]
 &  = \sum_{\ell = 0}^k \E\left[ \nabla_x J(x_i(n)) \cdot s^i(n) \middle| F_n, L = \ell \right]
        \pr\left[ L = \ell \right]  \\
 &  \geq \sum_{\ell = 0}^k \|\nabla_x J(x_i(n))\|_2\|s^i(n)\|_2 \left( \frac{1}{\sqrt2}(k - \ell) - \ell \right)
        2^\ell c^{\ell + 1}  \\
 &  = \frac{c\|\nabla_x J(x_i(n))\|_2\|s^i(n)\|_2}{\sqrt2}
        \left( \sum_{\ell = 0}^k (2c)^\ell
        \left( k - \left( 1 + \sqrt{2} \right) \ell \right) \right).
\end{align*}
It is easy to see that this quantity is positive for sufficiently large $k$,
since, as $k \to \infty$, $\sum_{\ell = 0}^k \ell (2c)^\ell$ converges while
$k\sum_{\ell = 0}^k (2c)^\ell$ diverges (recalling $c \in (0,1/4)$). \qed

{\bf Lemma 2}
The variance of the updates goes to zero as the cost function goes to zero.
That is, for some $K_0 > 0$,
\[\E\left[ \|s^i(n)\|^2 \right]
  \leq -K_0 \E\left[ \frac{dJ}{dx}(x_i(n))s^i(n) \right].\]

\emph{Proof of Lemma 2:}
We assume here that the food follows an isotropic Gaussian distribution
centered at the origin. That is, for some $\sigma > 0$, $\forall x \in \R^2$,
\[J(x)
  = \frac{1}{\sqrt{2\pi\sigma^2}}
    \exp\left(-\frac{\|x\|_2^2}{2\sigma^2} \right).\]
Then, for all agents $i$ and times $n$,
$$\frac{dJ}{dx}(x_i(n)) = -\frac{1}{\sqrt{2\pi}\sigma^2}
                \exp\left( -\frac{\|x_i(n)\|_2^2}{2\sigma ^2} \right)x_i(n)$$
Plugging in
$$x_i(n) = \sum_j a_{ij}(n)x_j(n)+\gamma_i(n)s^i(n)$$
gives
$$\frac{dJ}{dx}(x_i(n))=-\frac{1}{\sqrt{2\pi\sigma^2}}
  \exp\left(-\frac{\|x_i(n)\| ^2}{2\sigma^2}\right) \sum_j a_{ij}(n) x_j(n)
  - \frac{1}{\sqrt{2\pi\sigma}}
    \exp\left(-\frac{\|x_i(n)\| ^2}{2\sigma^2}\right) \gamma_i(n)s^i(n)$$
Now the right-hand side of the lemma:
\begin{align*}
& -K_0 \E \left[ \frac{dJ}{dx}(x_i(n))s^i(n) \right] \\
& = K_0 \E \left[ \frac{1}{\sqrt{2\pi \sigma^2}}
                  \exp \left( -\frac{\| x_i(n)\| ^2}{2\sigma^2} \right) s^i(n)
                  \sum_ja_{ij}(n)x_j(n)
                  + \frac{1}{\sqrt{2\pi\sigma^2}}
                    \exp \left( -\frac{\| x_i(n)\|^2}{2\sigma ^2} \right)
                    \gamma_i(n)s^i(n) \cdot s^i(n) \right] \\
& \geq K_0 \E \left[\frac{1}{\sqrt{2\pi\sigma^2}}
          \exp \left( -\frac{\| x_i(n)\| ^2}{2\sigma^2} \right)
          \gamma_i(n)\|s^i(n)\|^2 \right] \\
& = \E \left[ \exp \left( -\frac{\| x_i(n)\| ^2}{2\sigma ^2} \right)
            K'\gamma_i(n)\| s^i(n)\|^2 \right]
\geq \E \left[ \| s^i(n)\|^2 \right]
\end{align*}
when the step size $\gamma_i(n)$ satisfies
$K'\gamma_i(n) \geq \exp \left( \frac{\| x_i(n)\| ^2}{2\sigma ^2} \right)$.
\qed

\subsection*{Terrain Modeling and Swarm Initialization}
We use a similar terrain model to the one used in \cite{shklarsh}. Food density
and terrain are stationary over time. Food is assumed to be diffuse through the
terrain, with a global maximum at the source. Specifically, we modeled the food
density as an isotropic Gaussian function:
\begin{equation}
J(r)
  = \frac{K}{\sqrt{2\pi\sigma^2}} \exp \left( -\frac{r^2}{2\sigma^2} \right).
\label{eq:food_density}
\end{equation}
In addition, we introduce random obstacles in the form of local minima. In
particular, we use the half cosine function to generate field of obstacles:
\begin{equation*}
g(r)=min(0,-4(cos(\pi r/4)+0.5),
\end{equation*}
from which we randomly remove a small number of obstacles to increase
trial-to-trial variability.

\begin{figure}
\centering
\includegraphics[scale=.8]{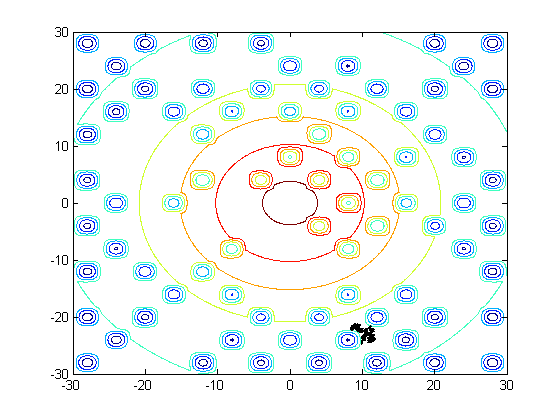}
\caption{A single swarm in our modelled terrain, immediately after
initialization. Contour lines illustrate food concentration (maximized at the
origin) and obstacles. Black dots indicate individual agents.}
\label{terrain}
\end{figure}

Each swarm is generated by choosing \emph{swarm center} at a certain fixed
distance from the food source, with the angle along this distance circle chosen
uniformly at random. Agents are placed uniformly in a small square centered at
the swarm center, with the constraint that no agent starts on an obstacle. See
Figure \ref{terrain} for an example of a single trial initialization.

\section*{Simulation parameters}
The results presented in the paper are produced with swarms consisting of 30
agents (unless otherwise stated). For the simulation we use a Repulsion Radius
of 0.1 and for the \cite{shklarsh} model we use 4.0 for Orientation Radius and
4.3 for the Attraction Radius. The discretized thresholding operator
$D_{L, T}$ in our model used $L = 4$ and $T = 3$ (see \cite{Emek}).

All the probability distributions are generated from $300$ independent trials.
For each trial the number of iterations required for at least $75\%$ of the
agents to reach within a fixed radius of the food source was measured. We have
set this source radius at $2.5$. For modeling the food source we have used
$\sigma^2 = 1000$ and $K = 200$ in equation (\ref{eq:food_density}).

\section*{Supporting Results}

\subsection*{Effect of different communication components}

In addition to the results presented in the main paper we also
tested the effects of each of the signals the agent / cell utilizes
as part of the DGD model. These include: i) repulsion information,
ii) orientation, and iii) attraction. We have evaluated the behavior
of the swarms in the absence of either of these component to
determine their impact of the ability of cells to efficiently reach
the food source.

\begin{figure}
\centering
\includegraphics[scale=.8,trim=0 5 0 0,clip]{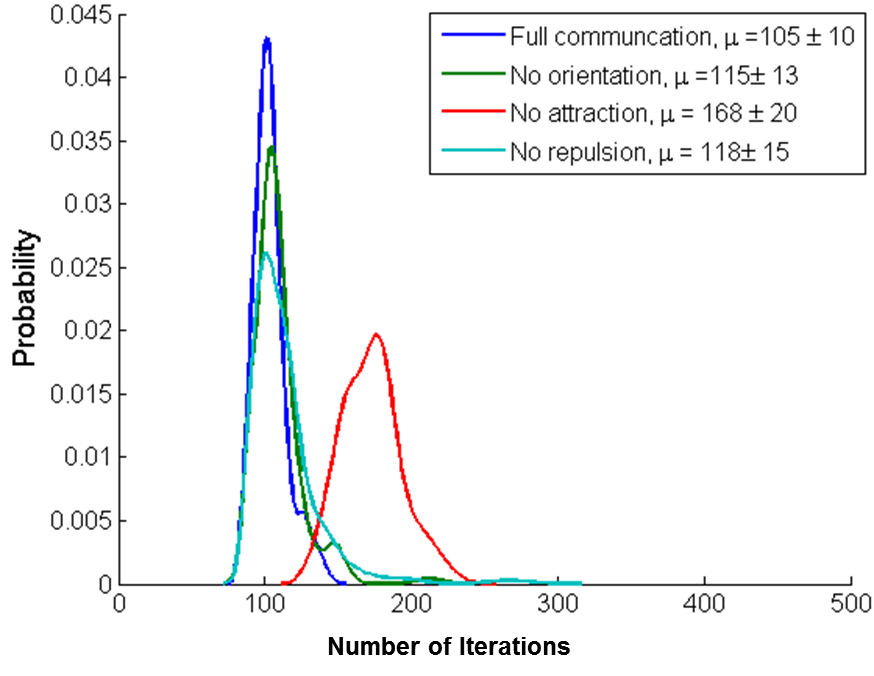}
\caption{Effect of the removal of different communication
components. $\mu$ denotes the average number of iterations required for each agent to reach the objective.}\label{comm}
\end{figure}

Figure ~\ref{comm} presents  the simulation results for this analysis.
As can be seen, while performance decreases when the orientation and
repulsion components are disabled (average number of iterations required to reach the food source increases from 105 to 118 for repulsion and 105 to 115 for orientation) the
effect is not large. In contrast the attraction component has a
large effect on performance. Specifically, removing this component
almost doubles the mean path length (168 iterations). One reason for this observation
is that the decay in the weights of attraction components is one fourth of
that of orientation (see methods section in the paper), providing
opportunity for the swarms to have a longer range of interactions
which dominates the swarm movement.

\subsection*{Effect of using distances to weight messages}

\begin{figure}
\centering
\includegraphics[scale=.8]{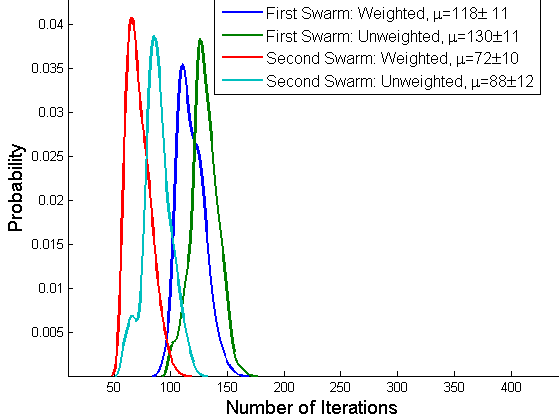}
\caption{Comparison of swarm performance using weighted and unweighted communication. $\mu$ denotes the average number of iterations required for each agent to reach the objective.}\label{weight}
\end{figure}

To test the impact of taking the distance into account when computing the local descent function we have run our method with and without using distance based weights, keeping all the other parameter settings same. Figure~\ref{weight} presents the results of this analysis using two sequential swarms. As can be seen, for both the first and second swarm, weighted communication leads to better performance than unweighted communication. Specifically, without using weights we see $10\%$ and $22\%$ increase in the number of iterations cells need to reach the food source for the first and second swarm, respectively. Thus, using the weighted version, which is also likely biologically correct, improves performance of such food search.
\subsection*{Effect of swarm size}
\begin{figure}
\centering
\includegraphics[scale=.65,trim=0 8 0 0,clip]{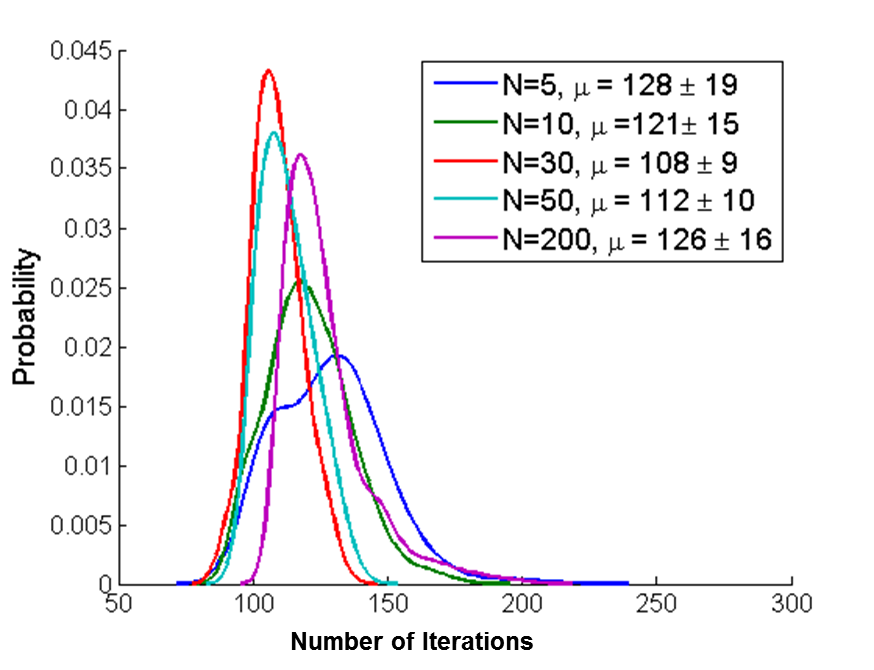}
\caption{Distribution of path length over different swarm sizes. $N$ denotes the number of agents in the swarm and $\mu$ denotes the average number of iterations required for each agent to reach the objective.}\label{food1}
\end{figure}

Another issue we tested is the impact of swarm size. As can be seen
in Figure~\ref{food1}), overall performance improves with the increase
in swarm size indicating that communication helps cells reach their
goal faster. However, depending on how we model the repulsion
distance, at some point such increase can lead to crowding. When the
agents are too close to each other, their movement is highly
constrained and mostly dominated by the repulsion effect. Therefore
it takes much longer for a larger swarm to reach the food source.
From our simulations we find that swarm size 30 to 50 gives the
optimal performance.

\subsection*{Model evaluation and comparison with previous experimental data}
To further test our model and compare it with the Shklarsh et al. model we used
experimental data from Taylor and Welch \cite{Taylor}. In their experiments,
Taylor and Welch focus on the effects of mobility on swarm performance when
a food gradient exists. Note that immobility of a large fraction of cells
directly affects swarms coordination since those that are still mobile are
receiving much fewer accurate messages impairing their ability to move in the
correct direction. Thus, as the fraction of immobile cells increases, we expect
several of the mobile cells to make at least some moves in the wrong direction.
Indeed, this is what Taylor and Welch observed in their experiments. To
quantify the impact of immobile agents they measured the ratio between cells
moving {\em towards} the food source and those moving {\em away} from the food
source at a specific time point. Formally, we define this as:
\begin{equation}
TR_i
 = \frac{\max \left\{ d^S-d_i^j: d_S>=d_i^j \right\}}
        {\left| \min \left\{ d^S-d_i^j: d^S<=d_i^j \right\} \right|}
\end{equation}
where $TR_i$ denotes the tracking ratio during iteration $i$, $d^S$ is the
initial distance of the swarm center from the food source and $d_i^j$ is the
distance of cell $j$ from the food source at iteration $i$. In their
experiments, cells  start far enough from the food source such that the impact
of the food gradient is minimal early on and  so cells depend on swarm
communication to move in the correct direction. Therefore, when a large fraction of
cells is immobile the swarm would expand in both directions, leading to a TR that
is close to 1. TR greater than 1 represents chemotactic (coordinated)
movement.

When mobility is drastically impaired (only 1\% of cells can move) Taylor and
Welch report a TR of close to 1. The TR increases as the fraction of
immobile cells decreases. These results were then used to identify mutants
associated with movement since their TR was reduced when compared to WT cells.

\begin{figure}
\centering
\vspace{-20pt}
\includegraphics[scale=.6]{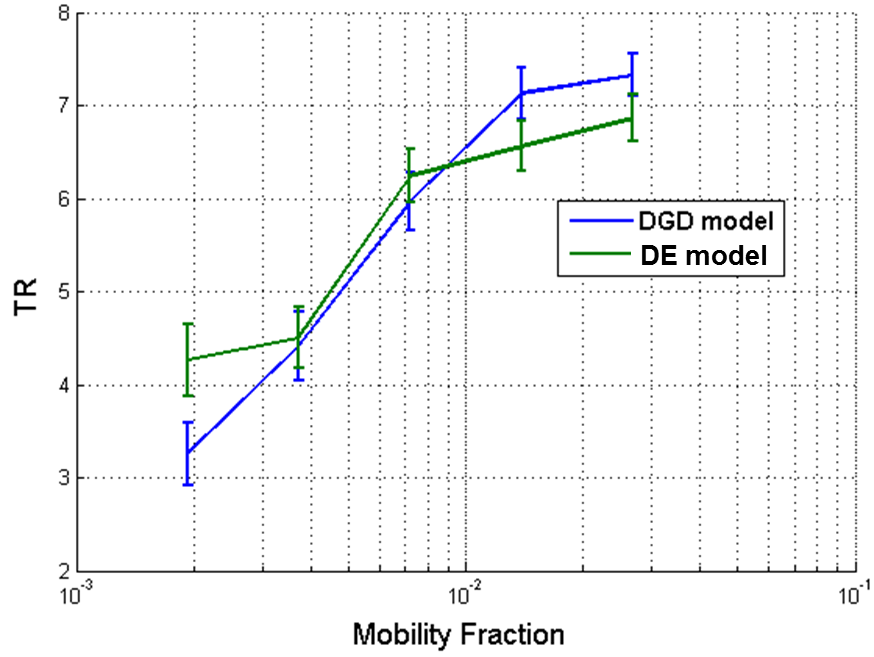}
\caption{Plot of TR vs mobility fraction. The results are generated with 300 independent trials using swarm population of 10000.}
\vspace{-25pt}
\label{TR}
\end{figure}

To test our model in this setting we have simulated the immobilization of
different fractions of cells and for each computed the resulting TR after a set
number of iterations for each fraction. Figure ~\ref{TR} presents the resulting tracking ratios when
using 10,000 cells. Note that while in our simulation we observe a TR that is
higher than 1, even when only 0.5\% are mobile, we attribute this to
the far smaller number of cells in our simulations. For such a small
set all mobile cells can likely communicate with all other mobile cells whereas
when the number of cells is larger this ability may be reduced. However, as was
observed in experimental results we also observe a rise in TR as the fraction of
mobile cells increases. Further, while the TR obtained by our DGD method for
the least constrained settings (3\% mobile) is slightly higher than the TR for the same setting according to Shklarsh et al., the initial TR is significantly lower
than the TR obtained by the Shklarsh model (and closer to the real result)
indicating that our model can still capture the communication capabilities of
bacterial cells with weaker assumptions and less computational power.
\begin{figure}
\centering
\includegraphics[scale=.35,trim=0 8 0 0,clip]{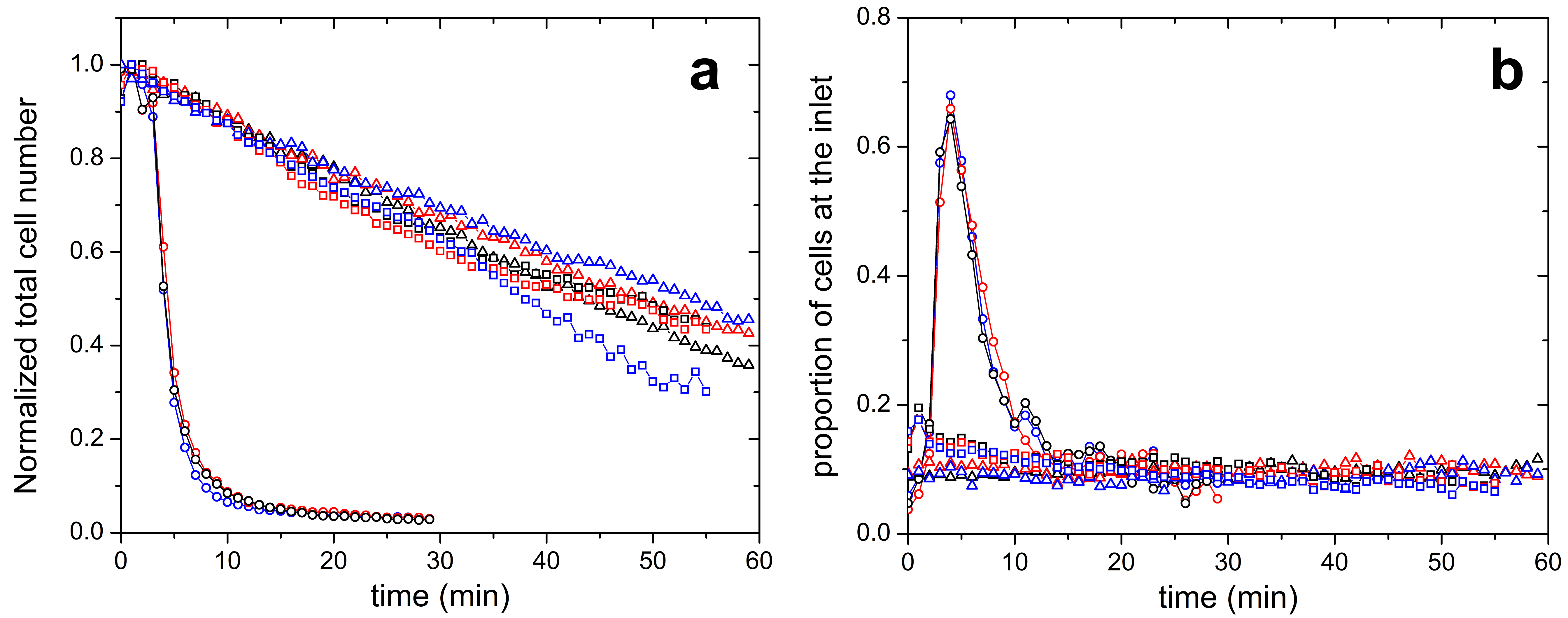}
\caption{The impact of communication on bacterial food search. Circles - WT bacteria. Rectangles - motility buffer, Triangles - $\Delta$ $tsr$. (a) Total cell number in the whole chamber. (b) The proportion of cells in the inlet area (377 μm radius from the inlet corner) as a function of time. Note that large increase in WT that is not observed in the communication-less conditions.}\label{Zhicheng_figure}
\end{figure}